%% file: main.tex
\documentclass[10pt, conference]{IEEEtran}
\IEEEoverridecommandlockouts
\usepackage{cite}
\usepackage{multirow}
\usepackage{amsmath,amssymb,amsfonts}
\usepackage{algorithm}
\usepackage{algpseudocode}
\usepackage{graphicx}
\usepackage{textcomp}
\usepackage{xcolor}
\usepackage{hyperref}
\usepackage{cleveref}

\newcommand{\etal}{\textit{et al.}}

\begin{document}

\title{Size matters? Or not: A/B testing with limited sample in automotive embedded software
\thanks{}
}

\author{\IEEEauthorblockN{Yuchu Liu}
\IEEEauthorblockA{
\textit{Volvo Cars}\\
Gothenburg, Sweden \\
yuchu.liu@volvocars.com}
\and
\IEEEauthorblockN{David Issa Mattos}
\IEEEauthorblockA{\textit{Computer Science and Engineering} \\
\textit{Chalmers University of Technology}\\
Gothenburg, Sweden \\
davidis@chalmers.se}
\and
\IEEEauthorblockN{Jan Bosch}
\IEEEauthorblockA{\textit{Computer Science and Engineering} \\
\textit{Chalmers University of Technology}\\
Gothenburg, Sweden \\
jan.bosch@chalmers.se}
\and
\IEEEauthorblockN{Helena Holmstr\"om Olsson}
\IEEEauthorblockA{\textit{Computer Science and Media Technology} \\
\textit{Malm\"o University}\\
Malm\"o, Sweden \\
helena.holmstrom.olsson@mau.se}
\and
\IEEEauthorblockN{Jonn Lantz}
\IEEEauthorblockA{
\textit{Volvo Cars}\\
Gothenburg, Sweden \\
jonn.lantz@volvocars.com}
}

\maketitle

\begin{abstract}
\input{sections/0abstract}
\end{abstract}


\input{sections/1introduction}
\input{sections/2background}
\input{sections/3method}
\input{sections/4BMW}
\input{sections/5casestudy}
\input{sections/6discussion}
\input{sections/7conclusion}




\section*{Acknowledgement}
This work is supported by Volvo Cars, by FFI, by the Wallenberg AI Autonomous Systems and Software Program (WASP) funded by the Knut and Alice Wallenberg Fundation and by the Software Center.
Furthermore, the authors would like to extend our gratitude to all participants of our case study, and to all members of the VEM functional development team.


\bibliographystyle{IEEEtran}
\bibliography{IEEEabrv, ref.bib}


\end{document}

%% file: sections/0abstract.tex
A/B testing is gaining attention in the automotive sector as a promising tool to measure causal effects from software changes. 
Different from the web-facing businesses, where A/B testing has been well-established, the automotive domain often suffers from limited eligible users to participate in online experiments.
To address this shortcoming, we present a method for designing balanced control and treatment groups so that sound conclusions can be drawn from experiments with considerably small sample sizes.
While the Balance Match Weighted method has been used in other domains such as medicine, this is the first paper to apply and evaluate it in the context of software development.
Furthermore, we describe the Balance Match Weighted method in detail and we conduct a case study together with an automotive manufacturer to apply the group design method in a fleet of vehicles.
Finally, we present our case study in the automotive software engineering domain, as well as a discussion on the benefits and limitations of the A/B group design method.

%% file: sections/1introduction.tex
\section{Introduction}
\label{intro}

A/B testing, or A/B experimentation, is an online experiment technique for evaluating causal effects from software changes \cite{Deng2013, Xie2016}.
In recent years, there is an increasing interest in adopting A/B testing in the automotive software businesses as it is considered an important tool for product development \cite{Giaimo2019, Mattos2020}.


As an online experimentation method, A/B testing relies on large sample sizes that are not always available in the automotive business.
With hundreds of millions of users, challenges in increasing sample size were experienced in the web domain as reported by \cite{Deng2013} and \cite{Xie2016}.
In the automotive domain, the available users to experiment with are notably more limited by orders of magnitude comparing to the web domain.
Since the most popular vehicles are sold in the ballpark of one hundred thousand units annually, with an average model sold in the range of tens of thousand units, and almost all vehicle models have local versions in their perspective sales markets.
Moreover, we would want to experiment with a fraction of the total population of vehicles. 
For instance, our experiments often only involve a particular model of a sedan with a specific electrical machine in Northern Europe, which further reduces the available sample sizes.
Thus, obtaining larger samples in automotive A/B testing is often unrealistic.

Two problems that occur from the use of experimentation with limited and small samples is the presence of random imbalance, i.e., the experimental groups are not compared prior to the experiment, and metric sensitiveness due to limited experimental power. 
Methods such as the CUPED (Controlled-experiment Using Pre-Experiment Data) and stratification \cite{Deng2013, Xie2016} are used in the web domain to increase the detection of changes within low sensitive metrics, i.e. metrics with high variance, however, they still require large sample sizes.
Moreover, these methods are commonly used only with a single covariate and cannot be used interchangeably with numerical and discrete covariates. 
Re-randomisation and seed selection are often a potential solution to random imbalance, but it can increase the time to conduct an experiment and they are not guaranteed to provide balanced groups if there are changes in the design of different experiments. Research literature on A/B testing, experimentation in the software domain and in the automotive software development does not provide clear guidance on how to conduct experiments with low sample size and potentially imbalanced groups.
Inspired by recent developments in the area of medicine and clinical trials, in this paper, we present a case study in automotive software utilising the Balance Match Weighted  method \cite{Xu2009} to create an experimental design that minimises group variance by balancing the control and the treatment groups with similar observed features (or, covariates). 
The design is guaranteed to provide maximum balance among the covariates and the analysis takes into account the covariates to reduce metric sensitiveness. 
Moreover, this design allows to include both numerical and categorical covariates.
In this paper, features and covariates refer to the independent variables in a statistical model and the two terms are used interchangeably.



The contribution of this paper is three-fold. 
First, we present the Balance Match Weighted method to design experiments in detail. While this design has been used in other areas of science, this is the first paper to apply it to experimental design in software development and in A/B testing.
Second, we provide a case study, in the automotive domain, of the Balance Match Weighted design. With this design, we are able to draw valid conclusions from the experiment with significantly lower sample sizes compared to the randomised field experiments usually conducted in the web domain. 
Third, we discuss the advantages and limitations of the  Balance Match Weighted in the design of experiments in the automotive domain.


The rest of this paper is arranged as follows. In Section \ref{related}, we present background and related work. Our research method is reported in Section \ref{method}. We describe the Balance Match Weighted design in detail in Section \ref{PSM}. The results from our empirical validation case study are presented in Section \ref{validation}. The discussion and conclusion are shown in Section \ref{discussion} Section and \ref{conclusion} respectively.

%% file: sections/2background.tex
\section{Background\label{related}}

The two-group design, also called A/B testing, is an experimental design method \cite{montgomery2017design, kohavi2009controlled}. In this design, users are randomly assigned to different variants of the product, the control variant (the current system) and the treatment variant (the system with a modification). The users are randomly assigned to different variants, and, after a period, the instrumented metrics for each variation are statistically compared.
One of the assumptions of this design is that if that the users of each variant group are equally comparable, i.e., the only systematic difference between them is the introduced software variant. 
If this assumption holds, the research and development organisation can establish a causal relationship between the software modification and the differences observed in the metrics. Kohavi \etal\ \cite{kohavi2009controlled} provide an in-depth discussion of common experimental design techniques used in online experiments.

Web-facing companies rely on randomisation and on a large number of users to ensure that the groups are comparable. 
However, due to the presence of random imbalance, even with large numbers randomisation is not guaranteed to produce comparable groups \cite{Gupta2018}. 
For instance, Bing used to re-randomise one out of four experiments due to random imbalance. 
Besides re-randomisation, Microsoft also utilises historical data to perform multiple A/A tests in order to find the best seeds to find the best-balanced groups \cite{Gupta2018}. 

A large number of diverse users in each experimental group lead to an increase in the variance of the metrics. 
This increase in the variance leads to less sensitive metrics \cite{Fabijan2017a}. 
Research on A/B testing has provided different statistical methods to reduce variance on experiments such as stratified sampling and the CUPED method (Controlled experiment Using Pre-Experiment Data) \cite{Deng2013, Xie2016}. 
The CUPED method is similar to using control covariates in a regression. 
This method utilises pre-experiment data to identify covariates that can reduce the variance in the estimation and compensates for it in the average treatment effect.

Our proposed approach using the Balanced Matched Weight method addresses both the balance of the groups in small samples as well as reduces the variance in the metrics. 
It requires pre-experiment data to identify the features (or covariates) to balance the groups and utilises these features in a regression framework to reduce the variance of the metrics similarly to the CUPED method.

%% file: sections/3method.tex
\section{Research method \label{method}}
The objective of this study is to explore and validate A/B group design with the Balance Match Weighted method, in order to effectively A/B test with limited samples in the automotive domain.
We employ a case study method to empirically explore the A/B group design method in an automotive company, following guidelines from \cite{Maxwell1992} and \cite{Walsham1995}.

This study is part of a larger research collaboration between several automotive companies aiming to introduce A/B testing at scale.
As a first step to adopt A/B testing, the study company has deployed fleets of vehicles driven by internal users as testbeds for developing software architectural solutions, data analytic tools, and so on.
Furthermore, as already identified in previous research \cite{Mattos2020, Giaimo2020, Mattos2018} one of the limitations of A/B testing in automotive companies is the smaller sample size. 
In this context, the goal of this research is to identify techniques and experimental design methods aimed at inference with small samples.

Within medicine, experimental design minimisation techniques are widely used for small samples experiments \cite{Xu2009, Stuart2010}, and for experimental designs where there is prior information regarding the user characteristics and large variances between the users \cite{Rubin2001}, the Balance Match Weighted yields good variance reduction by balancing the groups compared to full randomised experiments.

This paper investigates the use of the Balance Match Weighted for experimental design in the automotive domain. This is captured by the following research questions: 

\textbf{RQ1:} How can we apply Balance Match Weighted design for the partition of A/B groups in the automotive domain?

\textbf{RQ2:} What are the advantages and limitations of the Balance Match Weighted design in the automotive domain? 

To address these research questions, we conduct a case study with an automotive company.
The case study company is an automotive manufacturer. 
Their business includes the design, development, and manufacturing of passenger vehicles.
We chose to utilise a case study method for the following reasons.
First, A/B testing is not yet an adopted practice in the embedded system domain to the best of our knowledge, thus there is a lack of literature and empirical data.
Second, a well-designed case study allows us to empirically validate the method in an automotive context, and it allows us to analyse the advantages of the design in its intended applications.
 

\subsection{Data collection}

In our research, we take advantage of the resources in the case study company and utilise two main sources of data collection.
First, we actively worked with a software development team in situ for a period of six months.
This development team consists of 34 members, their roles include software engineer, product owner, data scientist, etc.
The team focuses on software solutions for vehicle energy management and optimisation. 
At least one of the authors participated in every project meeting, workshop, and discussions during the entire period, and provided inputs in relation to A/B testing.
We collected meeting notes and design documentation.

The second source of our data collection is quantitative.
From October 2020 to March 2021, we collect measurements from a fleet of 28 cars leased to employees of the case study company.
The vehicles are commissioned for acquiring immediate user feedback of novel functionalities and they are driven as regular private vehicles.
We instrument the vehicles with software that actively measures 51 signals through vehicle on-board sensors. 
The raw measurements are sent off-board and are permanently stored, and the research group is granted full access to the database.


\subsection{Validity considerations}
In this subsection, we present the threats to validity in our case study and how the threats are mitigated.
    \subsubsection{Internal validity}
The propensity score model in the design method makes a strong ignobility assumption, which assumes that the effect on the target variable from the unobserved variables is minimum.
Since our case study is designed around existing software, the observed variables are defined prior to the study.
This implies that some co-founding variables might not be observed.
No special action is taken to mitigate this risk as the ignobility assumption should be considered as an inherent limitation of the design method.

The quantitative trip data collection was done during a twenty-week period. 
We raised concerns on if an usage pattern can be established during a relativity short period.
We mitigate this risk in two ways.
First, after analysing the data, we have discovered that on the aggregated level, data from over 13,000 valid trips were collected and they are collected from a total of 205,000 kilometres driven distance.
On average, each vehicle has made more than 250 trips during the period.
Second, we have observed that the usage pattern of each individual vehicle does not differ drastically from week to week. 
Therefore, we consider the number of trip samples sufficient and we assume the seasonality effects in this fleet are low. 

    \subsubsection{External validity}
In this case study, we have applied the experiment design method to one software developed by one automotive company. 
We recognise the limitations of the approach and our findings might not be applicable to the entire automotive domain.
However, we believe the design method can be adapted to run A/B experiments on similar software developed by other automotive manufacturers, as we demonstrated the design method using quantitative usage data that is arguably independent of the vehicle manufacturer.

%% file: sections/4BMW.tex
\section{The Balance Match Weighted design \label{PSM}}

This section provides a detailed description of the theoretical background of the Balance Match Weighted design method. 

The Balance Match Weighted design is an extension to the propensity score matching method\cite{Rosenbaum1983}, proposed by Xu and Kalbfleisch \cite{Xu2009}.
In the original literature, the Balance Match Weighted design was used to select balanced groups for controlled experiments for medical research.
Similar to some medical applications, A/B experimentation in software testing is not a traditionally controlled experiment, i.e., we cannot manipulate the boundary conditions of when and how the software is used.
As a result, the measured treatment effects could be caused by other variables rather than the treatment itself.
These variables could also result in a large variation in the measured treatment response, thus making treatment effects undetectable.

Prior to the experiment, when treatment has not been applied and the outcome is not known, Balance Match Weighted can be used in pre-experiment data to select the participants for each experimental group \cite{Rubin2001, Xu2009, Stuart2010a}. 
After the experiments have been conducted, the covariates used in the Balance Match Weighted are used to reduce the variance in estimating the average treatment effect \cite{Stuart2010}.


\subsection{The Balance Match Weighted design}

Consider an A/B experiment, where the sample size $N$ is small. 
If the groups are partitioned at random, it is likely we produce unbalanced groups. 
As a result, the response measured in the target variable $Y$ could have high variance and we risk inconclusive experiments.

\begin{algorithm}
\caption{The Balance Match Weighted method}
\textbf{Inputs: } $M$ repetitions, $N$ total number of users participating in the experiment 
\begin{algorithmic}[1]
\State Identify the relevant features $\mathbf{X}$.
\State Randomise $N/2$ subjects in control ($\tau=0$) and another $N/2$ subjects in treatment ($\tau=1$) group.
\While{$m < M$}
     \State From the identified features $\mathbf{X}$, compute estimated propensity score distance $\delta k_n$.
     \State Perform greedy full matching based on the propensity score distance by minimizing $\Delta k_m = \sum ^{N/2}_{n=0} \delta k_n$.
     \State Record the triplet \{$\Delta k_m$, $n|\tau =1$, $n|\tau =0$\}  
\EndWhile
\State Select the control and treatment where $\Delta k_m$  is minimum.
\end{algorithmic}
\label{bmw_1}
\end{algorithm}

The Balance Match Weighed design was formulated by Xu and Kalbfleisch, the purpose is to reduce the imbalance and to increase the precision of the estimated treatment effect \cite{Xu2009}.
Comparing to the literature, we make a slight modification to the matching process to satisfy our constraints.
It is an iterative process, described as the algorithm in Algorithm \ref{bmw_1}. In the following subsections, we discuss each step of the algorithm.


\subsection{Feature selection}
\begin{figure}[t]
\centerline{\includegraphics[width=\linewidth]{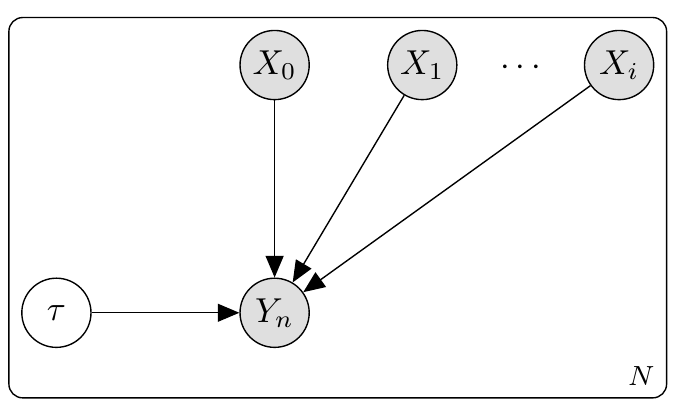}}
\caption{Relationships of input features ($\mathbf{X}$), treatment ($\tau$) and target variable ($Y$) in the propensity score matching model.}
\label{fig_pgm}
\end{figure}

We use a network diagram to illustrate relationships in features, the target variable, and the treatment, as shown in Fig. \ref{fig_pgm}.
The shaded nodes are observed variables, the transparent node indicates if the sample is in the control or treatment group, and arrows indicate dependency.

Consider a set of $i$ features, $\mathbf{X} = \{X_0, X_1, ..., X_i\} $, which are observed prior to the experiments. 
The observation is done for each individual subject sample $n$ in the entire sample set $N \in \{0, 1, ..., n\}$, and believed to be predictors to the target variable $Y$.
The changes in $Y$ are dependent on $\mathbf{X}$.
The target variable $Y$ is also what the treatment $\tau$ aims to influence.
We use $\tau$ as an indicator on whether the treatment is applied, $\tau \in \{0, 1\}$. $\mathbf{X}$ is independent from $\tau$.
We consider cases when control and treatment groups are even. 
Treatment will be applied to $N/2$ samples, with another $N/2$ in the control group.

In a successful A/B experiment, the treatment effect is sufficient to detect and therefore the expectation of the target variable $Y$ is, $E(Y|\mathbf{X}, \tau=1) - E(Y|\mathbf{X}, \tau=0) \neq 0$.

An important assumption made in the model is ignobility \cite{Rosenbaum1983, Stuart2010}. 
That is, we assume unobserved features do not affect the target variable $Y$.
To satisfy this assumption, an optimal model includes all known features which correlate to the target variable only and not to the treatment \cite{Rubin1996, Brookhart2006}.
As shown in Fig. \ref{fig_pgm}, there is no dependency in between features $\mathbf{X}$ and treatment $\tau$.
When the sample size $N$ is small, including a large number of $i$ features, might not be feasible \cite{Stuart2010}.
In this case, a recommendation made by Rubin \cite{Rubin2001} suggests to first include a small set of features known to be related to the target variable,
perform the matching and experiments, then include more features if bias is high in the outcome.
One should not include the target variable $Y$ in the propensity score model.

\subsection{Propensity score distance}

After selecting features that are highly informative of the target variable, the next step is to calculate the propensity score.
To compute the propensity score $\rho$, we fit the input features $\mathbf{X}$ to a logistic regression, with indicating variable $\tau = 0$ for the control group and $\tau = 1$ for the treatment group. 

We obtained the propensity score from the outcome of the logistic regression. The propensity score is a probability value that falls between 0 and 1. 
The individual propensity score distance for each subject $n$ is defined as the absolute difference of propensity score in the control $\left( \mathbf{X}|\tau=0 \right)$ and treatment $\left( \mathbf{X}|\tau=1 \right)$ group,

\begin{equation}
    \delta k_n = |\rho_{n, \tau=0} - \rho_{n, \tau=1}|
\end{equation}

where,

\begin{equation}
    \rho_{\tau=1} = P(Z=1|\mathbf{X}_n) = \frac{e^{\beta_{0} + \mathbf{\beta}\mathbf{X}}}{1 + e^{\beta_{0} + \mathbf{\beta}\mathbf{X}}}
\end{equation}

and $\mathbf{\beta}$ are fitted coefficients for the linear logistic regression model, $\beta_0$ is the fitted intercept. 
The total propensity score distance for all subjects in the control and treatment group is defined as:

\begin{equation}
    \Delta k =  \sum ^{N/2}_{n=0} \delta k_n
\end{equation}

Prior to an A/B experiment, the treatment indicator $\tau$ is unknown.
Therefore, in the scenario of calculating propensity scores to design experiment groups before the treatment is applied, we randomise the control and the treatment groups as step 2 of the Balance Match Weighted method.

\subsection{Greedy full matching}

After computing the propensity scores, one should perform a matching of control and treatment groups.
There are some commonly applied matching methods, including caliper matching, 1:1 nearest neighbour \cite{Stuart2010a} matching, and full matching \cite{Xu2009, Hansen2004}.

In the existing literature proposing the Balance Match Weighted design, matching is achieved through the optimal full match \cite{Hansen2004}.
Optimal full match makes replacement, meaning that one subject in the control group can be matched to multiple subjects in the treatment group.
Furthermore, optimal full match allows discarding of subjects from the sample group, which is considered as a hard constraint in our case study for the following two reasons.
First, our experiment subjects, the vehicles are costly to run without being included in the experiments.
Second, since the matching is done prior to the experiment with an unknown treatment effect, we do not yet know the target variable but an expected outcome, discarding subjects at this stage is considered premature.
Thus we suggest a greedy full matching should be performed to match all subjects. 
In practice, after the treatment is applied, one can discard subjects based on propensity scores computed from the actual control and treatment groups.

We formulate the matching of propensity scores as an optimisation problem, with the objective to minimise the global propensity score distance ($\Delta k$) in between control and treatment groups.
We perform the greedy matching without replacement as we assume all subjects are independent. 
This means that each subject in the control group can have only one corresponding subject in the treatment group. 

\subsection{The repetition parameter $M$}
    
\begin{figure}[t]
\centerline{\includegraphics[width=\linewidth]{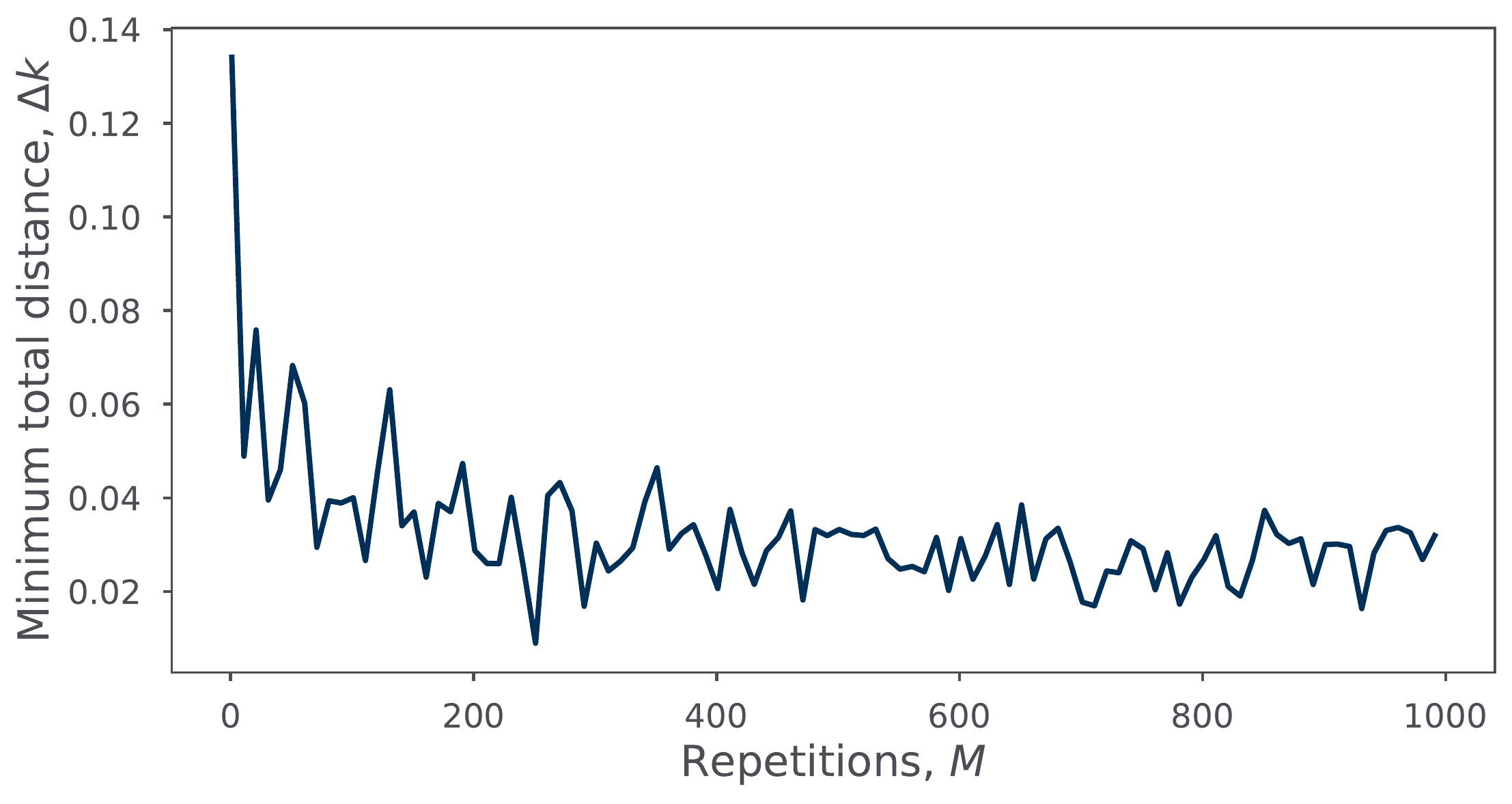}}
\caption{Minimum total distance ($\Delta k$) calculated from the propensity scores reduces as we increase the repetitions ($M$).}
\label{fig_bmw}
\end{figure}


In the literature, Xu and Kalbfleisch \cite{Xu2009} suggest the larger $M$ is, the better the results this design will obtain.
We decide on the number of repetitions by running the design with increasing repetitions of resolution 10, that is $M \in \{1, 10,..., 1000\}$, and analysing the trend of minimum total distance $\Delta k$ as $M$ increases.
We illustrate an example in Fig. \ref{fig_bmw}, for $N=28$ with six features, the improvement for $\Delta k$ becomes negligible after $M \approx 500$ repetitions.
An elbow effect is reached when the improvement of $\Delta k$ becomes minimum and the improvement no longer justifies the computational cost.

%% file: sections/5casestudy.tex
\section{Case study \label{validation}} 

We carried out our case study from October 2020 to March 2021, with a fleet of 28 passenger vehicles.
These vehicles belong to a large project of collaborative development with company car users from the case study company. 
All vehicles from the fleet are driven by corporate users as their primary family cars.
In our case study, we select vehicles of the same model with the same electrified propulsion, and the selected vehicle model is commercially available.
The vehicles are in the possession of users since December 2019, and all users reside in V\"{a}stra G\"{o}taland County, Sweden.

We aim to test the vehicle energy management (VEM) software, that can help reduce energy consumption through the prediction of vehicle routing.
From the confidential consideration of the software, we will not further discuss the functionality itself in this paper.
We introduce a variant A of the software to the fleet for a continuous period to collect pre-experiment data.
Utilising this data, we design the partitioning of control and treatment groups with the Balance Match Weighted method.
The B software variant is then shipped to the treatment group accordingly to the group partition.
The results of our group design with pre-experimental data, and the experiment design and outcome are presented in this section.

    \subsection{Case study fleet}

This case study is done in two distinct periods.
First, we have a twenty-week interval that is the observation period prior to applying treatment. 
Data generated from this period are used to partition the A/B groups. 
After that, there is an experiment period for two weeks during which the treatment is applied.
Data collected during the experiment period are for analysing the group design and the actual treatment effects.    

Aiming to simulate the real usage of cars in the case study, we do not dictate how the vehicles are driven. 
All the measurements and testing are done single-blinded, i.e., we do not interact with the users at all and they are not informed of the details of the A/B experiment.
Measurements are done through the on-board sensors of the vehicle.
We trust the measurements to a very large extent as the same sensors are used for calibration and diagnostic of all other functions in any commercially available vehicle.
The measurements are done continuously during all trips, and transmitted to a cloud storage through the telematics system of the vehicles via 4G.
The data generated are in time series at 10 Hertz, marked by an anonymous version of the vehicle identification number (VIN) for each vehicle and a unique ID for each trip.
We measure 51 signals from each vehicle, including but not limited to velocity, engine usage, climate system usage, GPS position, and so on.
In the data collection, we discard measured trips that are less than one minute in duration, or one hundred meters in distance.
In post-processing, we generate a number of observed features from the raw measurements. 

The VEM software tested in this paper was developed internally and shipped by the function development team.
The software was tested and validated through the standard processes in the case study company.
To enable full flexibility of an online A/B experiment, we adopt a hybrid architecture for this software.
The architecture determines that the software has two sets of parameters, local (A) and cloud (B). 
The local set of parameters are defaults for all vehicles with the same configuration.
While the cloud set of parameters are blank onboard but can receive external values from a cloud.
Since the VEM function is fully parameterised, this setup allows us to configure the function behaviour remotely.
During the observation period, all 28 vehicles are set on the A variant of parameters.
After the data is collected and analysed from the observation period, we partition the A and B groups and switch to the cloud parameters for group B.

    \subsection{Selected features}


We approach our feature selection both quantitatively and qualitatively.
As the vehicles were generating data from over 51 signals at 10 Hertz, on the weekly average, the dataset size is around one gigabyte when exported in the CSV format. 

In the quantitative selection process, we first aggregate all raw signals measured from the time series and compute the target variable $Y$  and all potential features from a few of these signals. We do this on both the trip level and car level.
We generate descriptive statistics to explore the correlation of target variables and all potential features.
In the end, we select six of the variables to be included in our features.
We examine the change of target variable over time, as well as its correlation to the features.
These variables are expected to be informative to the target variable.
The features are strongly correlated with the target variable and such correlation is consistent over time for the same vehicle.
Feature 0 and 1 have a negative correlation with the target variable, at -0.32 and -0.37 respectively.
Feature 2 through 5 are positively correlated with the target variable. The minimum correlation is at 0.26 in between feature 4 and the target variable, and the maximum correlation is 0.47 in between feature 5 and the target variable.

To ensure all known covariates which affect the target variable are included in the input features, we validate our features selected quantitatively with expert workshops.
We present our feature selection method and outcomes to a group of experts who actively developed the VEM software. 
Our proposal derived from data aligns with expert knowledge.
With that being said, we are testing a novel function which implies that we do not have more experience or data to rely on. 
We are aware there could be unobserved covariates that can affect the target variable.
Such shortcomings should be considered as an inherent limitation of the Balance Match Weighted design method for A/B experiments.

    \subsection{Matched A/B groups}

\begin{table*}[t]
\normalsize
\caption{Mean and variance of each of the five input features $\mathbf{X}$, min-max scaled, and propensity score in the matched control and treatment groups.}
\begin{center}
\begin{tabular}{|c|l|l|l|l|l|l|l|l|}
\hline
\multicolumn{1}{|l}{} &  & \textbf{Feature 0} & \textbf{Feature 1} & \textbf{Feature 2} & \textbf{Feature 3} & \textbf{Feature 4} & \textbf{Feature 5} & \textbf{Propensity score} \\ \hline \hline
\multirow{2}{*}{\textbf{Mean}} & Control & 0.51 & 0.64 & 0.46 & 0.42 & 0.53 & 0.48 & 0.49 \\ \cline{2-9} 
 & Treatment & 0.46 & 0.67 & 0.43 & 0.41 & 0.48 & 0.43 & 0.50 \\ \hline
\multirow{2}{*}{\textbf{Variance}} & Control & 0.06 & 0.03 & 0.10 & 0.06 & 0.08 & 0.06 & 0.000770 \\ \cline{2-9} 
 & Treatment & 0.04 & 0.06 & 0.07 & 0.07 & 0.08 & 0.08 & 0.000977 \\ \hline
\end{tabular}
\end{center}
\label{table_covar}
\end{table*}

At the end of the observation period, we collected the data from all vehicles. We extract the features $\mathbf{X}$ and the target variable $Y$ from the raw measurements.
The Balance Match Weighted design is applied to the dataset following the steps prescribed in Section \ref{PSM}. 

\begin{figure}[t]
\centerline{\includegraphics[width=\linewidth]{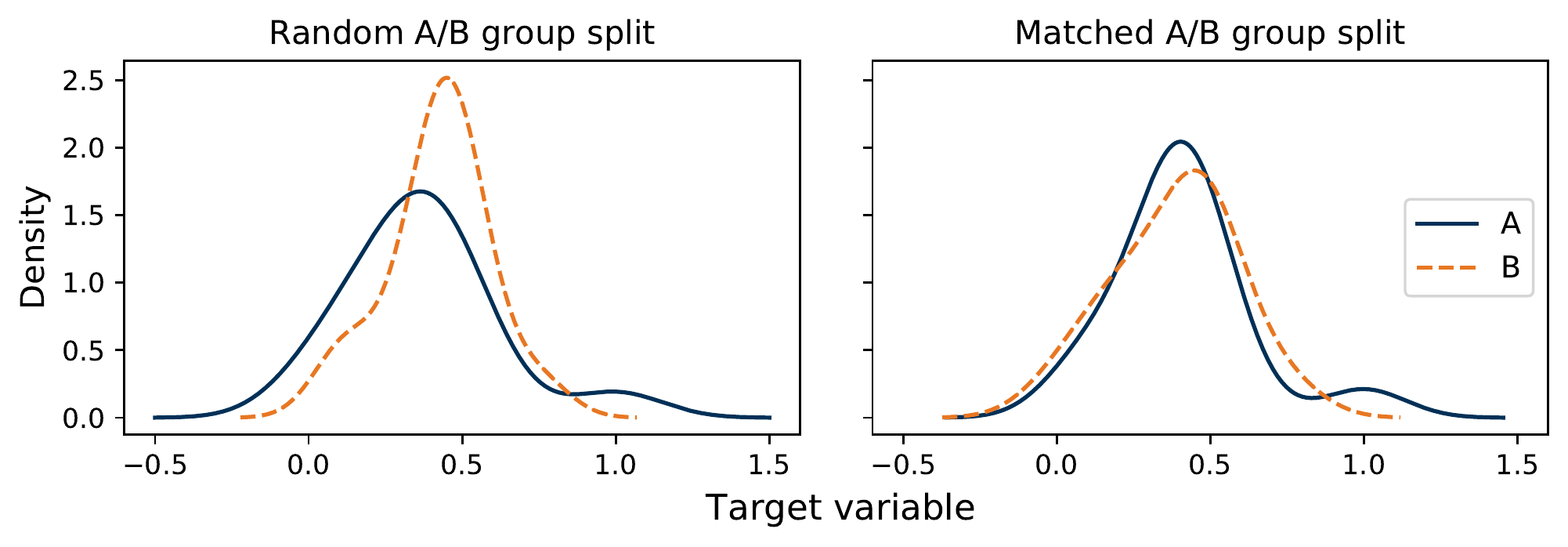}}
\caption{Kernel density estimation of the target variable, min-max scaled, of A and B groups when matched at random (left), and matched using the Balance Match Weighted design (right).}
\label{fig_target_kde}
\end{figure}

For each of the 28 subjects, the six features included in the model are aggregated on the vehicle level and stored in a $28 \times 6$ matrix.
That is, each vehicle has six features that represent its usage pattern, all of which strongly correlate to the target variable.
The target variable $Y$ is not included when estimating the propensity scores.
Before calculating the propensity score, the observed features are scaled with their perspective minimum and maximum values to minimise bias from extreme values in the observation.
We run the design with a high number of repetitions,  $M = 1000$, and obtain the A/B group partition with $N/2 = 14$ subjects in each group. 
We show the kernel density estimation of the target variable measured during the observation period. The distribution is shown for when the groups are partitioned are random, and when the groups are partitioned using the Balance Match Weighted design, see Fig. \ref{fig_target_kde}.
Comparing to random split, the matched A/B groups have a more balanced distribution of the target variable when the groups are running the same software.

The goal of matching is to achieve feature balancing, namely features in the control and treatment groups are from the same empirical distribution, $p\left( \mathbf{X}|\tau=0 \right)= p\left( \mathbf{X}|\tau=1\right)$.
Following advice from literature \cite{Rubin1996, Stuart2010}  , we further diagnose the validity of the matched groups by comparing their scaled mean and variance of the features.
We present the results in Table.\ref{table_covar}, as can be seen, the means are demonstrated to be similar in the matched groups as well as the variance in the two groups.
The average propensity score for the matched control and treatment group is 0.49 and 0.50, respectively, while the minimum values are 0.46 and 0.45, and the maximum values are 0.54 and 0.56, respectively.
The resulting group partition exhibits high similarities in the empirical distributions with merely 14 subjects in each group. 
A higher level of similarity in the empirical distributions would be expected, if the number of subjects $N$ increase \cite{Xu2009}.

    \subsection{Experiment outcome}

We conduct the A/B experimentation by introducing the B version of the software through a set of cloud-based parameters to the matched B group.
This A/B experiment serves as a demonstration of the Balance Match Weighted design method.
The experiment was run for a continuous two-week period, we measure the target variable $Y$ which is expected to reduce with using the new software.

We include an analysis of a paired test. A paired test is when we compare the target variable $Y$ from the same group of users, measured before and after the treatment is applied. 
This type of paired analysis can eliminate variation caused by the individual subject's preferences, however it is limited in quantifying external seasonality effects.
We include this analysis to illustrate the benefit of an A/B test with matched groups in comparison to a pseudo-random experiment. 
In the matched A/B test, the mean squared error (MSE) shows an 17.9\% improvement from the paired test, similarly to what is reported by \cite{Xu2009} at $N=30$ with eight input features.

In terms of the expected treatment effect $E(Y)$, We found that, comparing to a paired test, a matched group A/B test returns 37.87\% less standard deviation in the target variable.
The min-max scaled average treatment effect, $(\bar{Y}|\mathbf{X}, \tau=1) - (\bar{Y}|\mathbf{X}, \tau=0)$, is -0.134 and -0.180 for the paired test and matched A/B test, respectively.
The matched group A/B test yield lower variance in the target variable measured, at the same time returning a larger average treatment effect.


    \subsection{Recommended procedure}

We summarise the procedure of our experiment design and list them here in a step-wise manner.

\subsubsection{Determine eligible subjects and observe for a period}
The eligibility of the subjects shall be determined based on the purpose of the A/B tests. 
It is ideal if the subjects themselves are directly comparable.
For example, in the same A/B test, only the same vehicle model or engine types are included.
Or, if deemed necessary, the categorical variables or dummy variables which determine such vehicle properties can be included in the input features.
Moreover, decisions on the duration of data collection should take probable seasonality effects into consideration.
The seasonality effects could either be well-known before the observation starts or discovered during the observation. 
In the second case, the observation period should be reasonably extended to measure such effects.

\subsubsection{Select input features} 
The input feature selection shall be done both quantitatively and qualitatively.
As the data from the potential experiment subjects are collected through observations, selecting features based on pure statistical correlation might lead to spurious correlations and other problems alike.
We strongly recommend a more comprehensive approach that combines expert inputs with data, to ensure that all known covariates are taken into account in the model.

\subsubsection{Run the Balance Match Weighted design} 
For the case study, we have implemented the Balance Match Weighted design algorithm in Python.
The code takes the features $\mathbf{X}$, repetition number $M$ and total sample size $N$ as input, returns the ideal partition of A and B groups, the total propensity score distance $\Delta k$, and propensity score $\delta k_n$ for each pair of subjects.
The code will not be shared at this stage due to our confidentiality agreement with the case study company. 
However, there are similar and publicly available R packages\footnote{\href{https://github.com/markmfredrickson/optmatch}{https://github.com/markmfredrickson/optmatch}} for performing the matching design \cite{Hansen2004, Xu2009}.

\subsubsection{Apply treatment and collect data}
After the groups are designed and the treatment applied to the B group, the A/B experimentation shall run for a continuous period of time.
In our case study, we have predetermined the length of the experiment as we observed our subjects to have a consistent travel pattern from previous data.
To avoid false positive or false negative conclusions, if or when there is  a high variance of the features and the target variables over time, the experimentation shall not have a predetermined duration but terminate only when a conclusion is reached.

\subsubsection{Analyse experiment outcome}
When analysing the experimentation outcome, instead of directly computing the treatment response, one should also analyse if there is a significant difference in the input features before and after applying the treatment to validate the group partitioning.
This is to ensure the balanced group portioning modeled from pre-experimental data still holds, and the A/B groups are still directly comparable.
If there are discrepancies between the features measured prior and during the experimentation, one may perform a propensity score matching to only select subjects that are comparable.
Instead of comparing the point estimates, we suggest to visualise and compare the total distribution of the target variable measured in the A and the B groups. 

%% file: sections/6discussion.tex
\section{Discussion \label{discussion}}
In this section, we discuss the use, advantages, and limitations of the Balance Match Weighted design method applied in automotive software engineering.
This design method allows us to conduct A/B testing on limited samples, which is considered a major challenge in adopting A/B testing in the automotive domain \cite{Mattos2020}.
While our case study is an extreme example of limited sample size, through the study, we have demonstrated the intuitiveness of the group design method and the simplicity in adopting such method.
Small sample A/B testing can also be beneficial when applied in an agile development process, where the sample size can be gradually increased at each development iteration if the experiments are conclusive.
With that been said, we have discussed and experienced some limitations in applying the Balance Match Weighted design method in the automotive domain. 
They are listed in the subsections below.

\subsection{Existing data and unobserved variables}
Similar to the CUPED method \cite{Deng2013, Xie2016}, performing the Balance Match Weighted design requires pre-experimental data.
When the software in a novel product (e.g., a new model of vehicle) is the subject of interest, there might not be relevant existing data nor existing users.
The Balance Match Weighted design can only achieve balance in the features that are observed \cite{Rubin2001}. 
But when a novel software is being tested, we do not always have a comprehensive picture of which features should be included in the observation.
An incremental approach can be taken. To start, the development teams hypothesise the appropriate features and target variable prior, then gradually increase the number of features and sample size if the treatment effects are positive.
This development activity can be planned accordingly to agile methods.
As an alternative, an experiment group design method called Minimisation can be applied. Minimisation matches users as they enter the experiments \cite{Treasure1998}. 
Because it is reasonable to expect the number of eligible users to gradually increase.



\subsection{Multiple driver households and car sharing}
During the case study, we did not have any reliable method to define if/when the vehicle is driven by different people. 
Technology and privacy agreements limit the identification to individual vehicles only. 
This means that if there is more than one driver sharing a vehicle, we will not be able to capture the variation in the measurements caused by the driver change. 
However, we do not believe that this affects the case study outcome greatly as the VEM function does not interact directly with the drivers, therefore the function behaviour does not strongly depend on the preferences of individual drivers. 

In the automotive setting in general, we recognise the benefits of distinguishing multiple drivers sharing the same vehicle.
To capture the actual preferences of the drivers, user matching, partitioning of groups, and A/B testing should be done on the driver level instead of vehicles.
As driver distinction would generate more informative features, and an understanding of driver preferences is arguably necessary when user-facing software is tested.


%% file: sections/7conclusion.tex
\section{Conclusion \label{conclusion}}

A/B testing with limited samples is a challenge in the automotive sector.
To address this challenge, we evaluate and report an experiment group design method, Balance Weight Matched design, that can effectively increase the experiment power with small samples.
In this paper, we provide a detailed presentation of the design method and a step-by-step implementation procedure.
In collaboration with an automotive company, we conduct a case study to apply, demonstrate and evaluate the design method in a fleet of 28 vehicles with two versions of an energy optimisation software.
In the case study, we worked within a development team in situ.
From pre-experimental data, we found that feature balance can be achieved with merely 14 subjects in each group. 
After introducing the software treatment to the matched B group, compared to a paired test, the matched A/B test returns 37\% less standard deviations in the target variable while improving the MSE by 17\%.
We conclude that this design method is advantageous for conducting A/B testing in the automotive embedded software domain. 
As shown in our case study, balanced groups can be produced when the sample sizes are considerably small and it improves the power of small sample experiments.
Finally, we discuss some potential challenges and limitations in applying the Balance Weight Matched design in the automotive domain, including the ignorability assumption, conducting experiments with no prior data and we highlight the importance of differentiating drivers when a vehicle is shared.

In our future work, we plan to further investigate the Balance Weight Matched design method with more datasets and software, as well as develop tools for experiment analysis.

%% file: main.bbl
\begin{thebibliography}{10}
\providecommand{\url}[1]{#1}
\csname url@samestyle\endcsname
\providecommand{\newblock}{\relax}
\providecommand{\bibinfo}[2]{#2}
\providecommand{\BIBentrySTDinterwordspacing}{\spaceskip=0pt\relax}
\providecommand{\BIBentryALTinterwordstretchfactor}{4}
\providecommand{\BIBentryALTinterwordspacing}{\spaceskip=\fontdimen2\font plus
\BIBentryALTinterwordstretchfactor\fontdimen3\font minus
  \fontdimen4\font\relax}
\providecommand{\BIBforeignlanguage}[2]{{%
\expandafter\ifx\csname l@#1\endcsname\relax
\typeout{** WARNING: IEEEtran.bst: No hyphenation pattern has been}%
\typeout{** loaded for the language `#1'. Using the pattern for}%
\typeout{** the default language instead.}%
\else
\language=\csname l@#1\endcsname
\fi
#2}}
\providecommand{\BIBdecl}{\relax}
\BIBdecl

\bibitem{Deng2013}
A.~Deng, Y.~Xu, R.~Kohavi, and T.~Walker, ``Improving the sensitivity of online
  controlled experiments by utilizing pre-experiment data,'' in
  \emph{Proceedings of the sixth {ACM} international conference on Web search
  and data mining - {WSDM} {\textquotesingle}13}.\hskip 1em plus 0.5em minus
  0.4em\relax {ACM} Press, 2013.

\bibitem{Xie2016}
H.~Xie and J.~Aurisset, ``Improving the sensitivity of online controlled
  experiments,'' in \emph{Proceedings of the 22nd {ACM} {SIGKDD} International
  Conference on Knowledge Discovery and Data Mining}.\hskip 1em plus 0.5em
  minus 0.4em\relax {ACM}, aug 2016.

\bibitem{Giaimo2019}
F.~Giaimo, H.~Andrade, and C.~Berger, ``The automotive take on continuous
  experimentation: A multiple case study,'' in \emph{2019 45th Euromicro
  Conference on Software Engineering and Advanced Applications ({SEAA})}.\hskip
  1em plus 0.5em minus 0.4em\relax {IEEE}, aug 2019.

\bibitem{Mattos2020}
D.~I. Mattos, J.~Bosch, H.~H. Olsson, A.~M. Korshani, and J.~Lantz,
  ``Automotive {A}/{B} testing: Challenges and lessons learned from practice,''
  in \emph{2020 46th Euromicro Conference on Software Engineering and Advanced
  Applications ({SEAA})}.\hskip 1em plus 0.5em minus 0.4em\relax {IEEE}, aug
  2020.

\bibitem{Xu2009}
Z.~Xu and J.~D. Kalbfleisch, ``Propensity score matching in randomized clinical
  trials,'' \emph{Biometrics}, vol.~66, no.~3, pp. 813--823, nov 2009.

\bibitem{montgomery2017design}
D.~C. Montgomery, \emph{Design and analysis of experiments}.\hskip 1em plus
  0.5em minus 0.4em\relax John wiley \& sons, 2017.

\bibitem{kohavi2009controlled}
R.~Kohavi, R.~Longbotham, D.~Sommerfield, and R.~M. Henne, ``Controlled
  experiments on the web: survey and practical guide,'' \emph{Data mining and
  knowledge discovery}, vol.~18, no.~1, pp. 140--181, 2009.

\bibitem{Gupta2018}
S.~Gupta, L.~Ulanova, S.~Bhardwaj, P.~Dmitriev, P.~Raff, and A.~Fabijan, ``The
  anatomy of a large-scale experimentation platform,'' in \emph{2018 {IEEE}
  International Conference on Software Architecture ({ICSA})}.\hskip 1em plus
  0.5em minus 0.4em\relax {IEEE}, apr 2018.

\bibitem{Fabijan2017a}
A.~Fabijan, P.~Dmitriev, H.~H. Olsson, and J.~Bosch, ``The benefits of
  controlled experimentation at scale,'' in \emph{2017 43rd Euromicro
  Conference on Software Engineering and Advanced Applications ({SEAA})}.\hskip
  1em plus 0.5em minus 0.4em\relax {IEEE}, aug 2017.

\bibitem{Maxwell1992}
J.~Maxwell, ``Understanding and validity in qualitative research,''
  \emph{Harvard Educational Review}, vol.~62, no.~3, pp. 279--301, sep 1992.

\bibitem{Walsham1995}
G.~Walsham, ``Interpretive case studies in {IS} research: nature and method,''
  \emph{European Journal of Information Systems}, vol.~4, no.~2, pp. 74--81,
  may 1995.

\bibitem{Giaimo2020}
F.~Giaimo, H.~Andrade, and C.~Berger, ``Continuous experimentation and the
  cyber{\textendash}physical systems challenge: An overview of the literature
  and the industrial perspective,'' \emph{Journal of Systems and Software},
  vol. 170, p. 110781, dec 2020.

\bibitem{Mattos2018}
D.~I. Mattos, J.~Bosch, and H.~H. Olsson, ``Challenges and strategies for
  undertaking continuous experimentation to embedded systems: Industry and
  research perspectives,'' in \emph{Lecture Notes in Business Information
  Processing}.\hskip 1em plus 0.5em minus 0.4em\relax Springer International
  Publishing, 2018, pp. 277--292.

\bibitem{Stuart2010}
E.~A. Stuart, ``Matching methods for causal inference: A review and a look
  forward,'' \emph{Statistical Science}, vol.~25, no.~1, pp. 1--21, feb 2010.

\bibitem{Rubin2001}
D.~B. Rubin, ``Using propensity scores to help design observational studies:
  Application to the tobacco litigation,'' \emph{Health Services and Outcomes
  Research Methodology}, vol.~2, no. 3/4, pp. 169--188, 2001.

\bibitem{Rosenbaum1983}
P.~R. Rosenbaum and D.~B. Rubin, ``The central role of the propensity score in
  observational studies for causal effects,'' \emph{Biometrika}, vol.~70,
  no.~1, pp. 41--55, 1983.

\bibitem{Stuart2010a}
E.~A. Stuart and N.~S. Lalongo, ``Matching methods for selection of
  participants for follow-up,'' \emph{Multivariate Behavioral Research},
  vol.~45, no.~4, pp. 746--765, aug 2010.

\bibitem{Rubin1996}
D.~B. Rubin and N.~Thomas, ``Matching using estimated propensity scores:
  Relating theory to practice,'' \emph{Biometrics}, vol.~52, no.~1, p. 249, mar
  1996.

\bibitem{Brookhart2006}
M.~A. Brookhart, S.~Schneeweiss, K.~J. Rothman, R.~J. Glynn, J.~Avorn, and
  T.~Stürmer, ``Variable selection for propensity score models,''
  \emph{American Journal of Epidemiology}, vol. 163, no.~12, pp. 1149--1156,
  apr 2006.

\bibitem{Hansen2004}
B.~B. Hansen, ``Full matching in an observational study of coaching for the
  {SAT},'' \emph{Journal of the American Statistical Association}, vol.~99, no.
  467, pp. 609--618, sep 2004.

\bibitem{Treasure1998}
T.~Treasure and K.~D. MacRae, ``Minimisation: the platinum standard for
  trials?'' \emph{{BMJ}}, vol. 317, no. 7155, pp. 362--363, aug 1998.

\end{thebibliography}
